# Polemical Case Study of Opinion Dynamics: Patterns of Filter Bubbles in Non-Consensus, Rewire Phenomena

Yasuko Kawahata [‡]

[‡] Faculty of Sociology, Department of Media Sociology, Rikkyo University, 3-34-1 Nishi-Ikebukuro, Toshima-ku, Tokyo, 171-8501, JAPAN.
ykawahata@rikkyo.ac.jp, kawahata.lab3@damp.tottori-u.ac.jp

**Abstract:** In this paper, we will review some of the issues that have been raised by opinion dynamics theory to date. In particular, we conducted a hypothesis-based simulation using a socio-physical approach regarding the filter bubble phenomenon that tends to occur under special conditions such as (1) Distance, (2) Time, and (3) Existence of strong opinion clusters in the barriers to consensus building (4) Place where opinions are not influenced In particular, this paper discusses the hypothesis and simulations of filter bubbles, in which opinions diverge or converge without reaching consensus under conditions in which non-consensus is likely to be emphasized. This article serves as an Appendix in "Case Study On Opinion Dynamics In Dyadic Risk On Possibilities Of Trust-Distrust Model."

## 1. Introduction

The study of opinion dynamics has a long history and has been the subject of much research, mainly in the field of sociology. Early studies assumed linearity, but models incorporating nonlinearity were also studied. Consensus formation based on local majority rule has been studied as an application of renormalization group theory in physics. Also, a theory that compares the agreement and disagreement of opinions with the direction of the magnetic moment of magnetism has been studied in the field of social physics by applying the theory of magnetic physics by Garam et al. (1982). Many mathematical theories of opinion dynamics treat opinions as discrete values of +1 and 0 or +1 and -1. In contrast, some theories consider opinions as continuous values, which can be varied through the exchange of opinions with others. The bounded-confidence model is a typical model of a theory that deals with continuous transitions of opinions. The proliferation of public networks has enabled instantaneous two-way communication that transcends temporal and spatial constraints. The vast amount of textual data on the Web facilitates quantitative analytical studies of public opinion, which could not be visualized before. In this paper, we review the issues raised by previous opinion dynamics theories. In particular, we conduct simulations based on hypotheses from a socio-physical approach on the filter bubble phenomenon, which tends to occur under special conditions such as below.

(1) Distance

(2) Time

(3) The existence of Strong Opinion

(4) Place where Opinions are not Influenced

These clusters that act as barriers to consensus formation. Therefore, we are working on a theory to explain consensus formation and opinion splitting in opinion exchanges on social media such as Twitter (X). In this paper, we propose a model based on the Like Bounded Confidence Model, which represents opinions as continuous quantity values. However, the Bounded Confidence $m$Model assumed that people with differing opinions work by ignoring, rather than ignoring, their opinions, but in this paper, the authors' approach, especially when opinions are strong or when filter bubbles or echo chambers occur, Hypotheses and considerations are presented, and the model is designed to incorporate and represent the effects of external external pressures and phenomena that depend on the surrounding circumstances.

### 1.1 Research Focus

In particular, this paper addresses the filter bubble phenomenon. In particular, the above phenomenon occurs in situations where non-consensus is likely to be emphasized. In this case, the authors hypothesize that opinions tend to diverge or converge without reaching consensus. The filter bubble in this case is discussed in terms of hypotheses and simulations. And this article serves as an Appendix in "Case Study On Opinion Dynamics In Dyadic Risk On Possibilities Of Trust-Distrust Model."

## 2. Preview Works about Filter bubble

Filter bubbles are defined as the individual outcomes of different processes of information retrieval, perception, and selection, and by remembering the sum total, the individual user receives only customized choices from the world of available information that fit his or her existing attitudes. At the social level, individuals tend to share a common social media bubble with like-minded friends (Boutyline and Willer, 2017 ; McPherson, Smith-Lovin, and Cook, 2001). Similarly, the definition of an echo chamber has been said that over time, communities in which Internet content confirming a particular ideology is echoed by all sides are particularly prone to a process of group radicalization and polarization (Vinokur and Burnstein, 1978, Garrett, 2009 ; Sunstein, 2001, 2009). By this, diversity can be understood as source or content diversity. Source diversity refers to both the inclusion of a large number of information sources in a news outlet and the inclusion of a variety of referents in a news article. A wide range of different areas of interest in a particular topic, as well as in the selection of viewpoints offered to news consumers throughout, yields a diversity of content (Voakes et al. 1996) is also assumed. Scholars have expressed concern about whether algorithms value diversity as an important feature of news quality (Pasquale 2015). Theoretical concepts such as Pariser's (2011) filter bubble hypothesis suggest that algorithms aim to maximize economic benefits by increasing media consumption rather than guaranteeing diversity. According to this rationale, the algorithm excludes information that appears to be of little interest to individual users while presenting more content that they are more likely to consume. For example, a user who has experienced heavy sports news consumption will likely receive more sports news at the expense of other topics (e.g., political news).

### 2.1 Echo chambers

One might argue that the creation of "echo chambers" is also possible in the offline world simply by consuming certain television channels or newspapers. However, an increasing number of studies have recently hypothesized that creating "echo chambers" on the Internet is easy. Echo chambers are social phenomena in which the filter bubbles of interacting individuals overlap strongly. The danger of a society collapsing into distinct echo chambers could be explained as a lack of consensus throughout society, and a lack of at least some shared beliefs among otherwise disagreeing people, necessary for a democratic decision-making process (Sunstein, 2001, 2009 ). However, increasingly radicalized ideological online groups may, at some point, resort to real violence or terrorism to achieve their goals (e.g., Holtz, Wagner, and Sartawi, 2015 ; Weiman , 2006). For example, the Internet is an environment of choice, offering the possibility to meet many individuals around the world and breaking down regional limitations. In conclusion, it appears that users of social media platforms and social networking sites are at risk of both "filter bubbles" and "echo chambers". Similar theoretical constructs aim to increase the likelihood of like-minded contacts ("echo chambers"; Sunstein 2009) and a limited public sphere ("spheres"; Gitlin 1998). The latter, in particular, refers to the normative fear of unintentionally missing out on a variety of information that prevents individuals from being properly informed and becoming rational democratic citizens.

They report that a system of pre-selection/implicit selection of personalized information may actually lead to a reduction in the presentation and consumption of anti-attitudinal information (Beam, 2014). Furthermore, one study found that approximately 12% of Google web search results show differences between users. This can be explained by pre-selected/implicit personalization (Hannak et al., 2013). Both studies support the "filter bubble" hypothesis. Another study showed that individuals actually choose to read news items that seem consistent with their opinions (Garrett, 2009). However, the effect on avoidance of anti-attitude news items was reported to be less pronounced in this study (Garrett, 2009). Furthermore, studies by Iyengar and Hahn (2009) and Peterson etal (2018) indicate that individuals prefer to read news articles, news websites, and content that align with their political orientation. Their findings further underscore the "echo chamber" hypothesis.

They diagnose a high level of uncertainty about privacy issues on the Internet. Users are usually unaware of how their data will be used. Even those who express privacy concerns may be providing sensitive information due to a lack of awareness of the issue. Ideally, users who are concerned about privacy would want to control what information is shared and with whom. In practice, however, companies have ways to get people to share, such as creating default settings on their sites or giving the impression that everyone else is also sharing information. In fact, we have become accustomed to the idea that handing over personal data is the price you pay for a free service, that extra convenience [1].

### 2.2 Move on to Case study Modeling:Fillter Bubbles

In this paper, we hypothesize and simulate the spread of this "non-consensual" information, the filter bubble. In an actual research case study on the filter bubble, an exploratory study of COVID-19 misinformation on Twitter (2021) found that by July 18, 2020, the International Fact-Checking Network (IFCN), which integrates over 92 fact-checking organizations, had identified 7,623 pandemic-related IFCN has uncovered more than 7,623 unique fact-checked articles on the pandemic. But misinformation does more than contribute to the spread. Misinformation intensified fear, caused social discord, and could lead to direct damage. (e.g., deliberately engaging in dangerous behavior) .

## 3. Approrch to Fillter Bubbles Cases by socio-physics

The filter bubble hypothesized in this paper is the "filter bubble phenomenon" when a topic already existing in a certain space, in another discourse, is re-looped within a certain community, or when the case of a topic propagating to a completely unrelated community is repeated. Here, we will hypothesize the process of topic re-propagation and repetition in terms of topic re-wiring to a certain agent.

### Rewiring Process

The rewiring process is a stochastic procedure where edges between nodes may be reconfigured. The process is conducted through the following steps:

(1) Random selection of nodes based on opinion threshold and rewiring probability.
(2) Edges are reconfigured among the selected nodes to alter the information flow in the network.

This process can be represented as a transition in edge configuration from $E$ to $E'$.

### Opinion Formation

Opinion formation is modeled as an iterative process where each node updates its opinion based on the average opinions of its neighbors:

$$o_i(t+1) = \frac{1}{|N_i|} \sum_{j \in N_i} o_j(t), \quad (1)$$

where $N_i$ represents the set of neighbors of node $i$, and $t$ indicates the iteration step.

### Distribution of distances changes after each rewiring step

To express how the distribution of distances changes after each rewiring step, we use the notion of distance as a random variable. Let $D_{ij}^{(s)}$ be the distance from node $i$ to node $j$ at step $s$. The set of distances between all nodes is considered to follow the following probability distribution:

$$P(D_{ij}^{(s)} = d) = \frac{\text{step } s \text{ distances } d \text{ Pairs of Node}}{\text{all Nodes}} \quad (2)$$

where $P(D_{ij}^{(s)} = d)$ is the probability that the distance between two nodes is $d$ at step $s$.

The average value of the distance is expressed in terms of expectation, and the average distance for all node pairs at a particular rewiring step $s$ can be obtained:

$$E[D^{(s)}] = \sum_{i \neq j} D_{ij}^{(s)} \cdot P(D_{ij}^{(s)} = d) \quad (3)$$

This expected value $E[D^{(s)}]$ gives the average distance between nodes at step $s$. As the dynamics of the network progresses, these values will show changes that reflect patterns of information transfer and connectivity within the network.

### Conditional Node Selection in Network Rewiring

During the rewiring process, nodes are selectively subjected to the possibility of rewiring based on a stochastic condition influenced by their opinion values. This selection process is governed by the following probabilistic rule:

$$\text{select } node \text{ with probability } P(\text{select} \mid o_{\text{node}}) = \begin{cases} 1 & \text{if } o_{\text{node}} \leq t, \\ 0.5 & \text{if } o_{\text{node}} > t. \end{cases}$$

Where:

$t$ represents a threshold value for opinions. The probability with which a node is selected is determined based on this threshold in relation to the node's opinion value $o_{\text{node}}$.

If the opinion $o_{\text{node}}$ of a particular node is less than or equal to $t$, then that node is selected with a probability of 1.

Conversely, if $o_{\text{node}}$ is greater than $t$, the node is selected with a probability of 0.5.

Here, $o_{\text{node}}$ represents the opinion value of the node, and $P(\text{select} \mid o_{\text{node}})$ is the conditional probability of selecting the node for potential rewiring given its opinion value. The process aims at randomly choosing nodes, biased towards those with opinions equal to or less than 0.5. Nodes are added to the set of selected nodes until the set reaches a predetermined size, in this case, four nodes.

### Opinion Formation Process

The opinion formation process in the network is an iterative procedure where the opinion of each node is influenced by the opinions of its predecessors. This is mathematically represented by the following equation and is executed iteratively for a specified number of steps or until the system reaches a steady state.

For each iteration $t$, the opinion $o_i(t)$ of each node $i$ is updated based on the opinions of its predecessor nodes. The updated opinion $o_i(t+1)$ is calculated as follows:

$$o_i(t+1) = \frac{1}{|N_i|} \sum_{j \in N_i} o_j(t), \qquad (4)$$

where:

$o_i(t+1)$ is the opinion of node $i$ at iteration $t+1$,

$N_i$ is the set of predecessors of node $i$ in the network,

$|N_i|$ is the number of nodes in $N_i$,

$o_j(t)$ is the opinion of node $j$ at iteration $t$, and node $j$ is a member of $N_i$.

The process is repeated for a number of iterations, or until the opinions in the network stabilize, leading to the final opinion values for each node in the network.

### Model Output

The final state of the system is represented by:

(1) The adjusted opinion vector $\mathbf{o}' = (o'_1, o'_2, \ldots, o'_N)$ after the opinion formation process.
(2) A visualization of the network, indicating the opinion values through node coloring.
(3) A histogram representing the distribution of rewirings across nodes.

The model involves parameters such as the rewiring probability and opinion dynamics that can be varied to study different scenarios of social influence and information flow within a network.

Each node in the network represents an individual agent (an individual with an opinion), and the links between agents indicate the possibility of exchanging opinions.

(1) Generating nodes with random opinion values
(2) Add random directed link
(3) Forming filter bubbles by rewiring links
(4) Opinion formation process

### Cases of falling into a filter bubble Formula

Opinion updating is done by averaging the opinions of a node's predecessor nodes:

$$o_i^{(t+1)} = \frac{1}{N_{\text{neighbors}}} \sum_{j \in \text{neighbors}(i)} o_j^{(t)}$$

Here, $o_i^{(t)}$ is the opinion of node $i$ at time $t$, and neighbors($i$) is the set of precursor nodes of node $i$. represents.

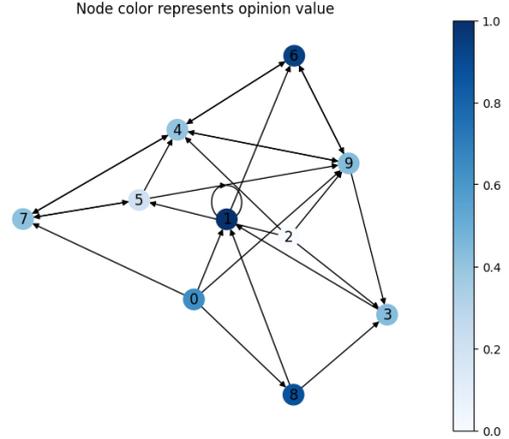

Fig. 1: falling into a filter bubble during the consensus building process $N = 10$

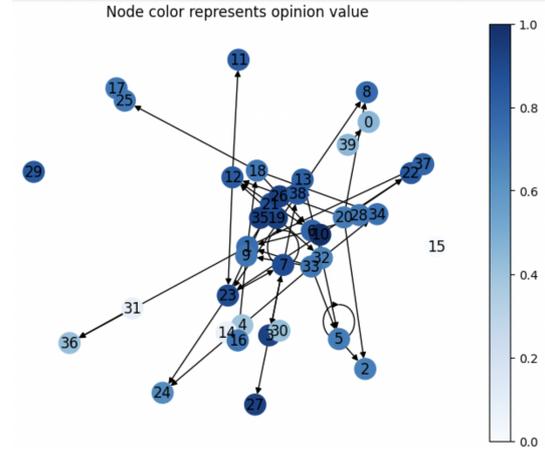

Fig. 2: falling into a filter bubble during the consensus building process $N = 40$

### Reproducing the filter bubble

When the opinion value exceeds a certain threshold, the node stops updating its opinion. Because it is meant to reduce diversity of opinion and create filter bubbles:

$$o_i^{(t+1)} = \begin{cases} \frac{1}{N_{\text{neighbors}}} \sum_{j \in \text{neighbors}(i)} o_j^{(t)} & \text{if } o_i^{(t)} < \text{threshold} \\ o_i^{(t)} & \text{otherwise} \end{cases}$$

### Parameter

The following key parameters are used in the simulation:

$N$: Number of nodes in the network (e.g. $N = 10$, $N = 40$)

threshold: Threshold of opinion value to stop updating opinions (e.g. threshold = 0.8)

Figure 1, Figure 2 shows,

(1) **Tendency of Agent Rewiring**

Some nodes (agents) have a large number of ingress and egress links, while other nodes have very few links. This suggests that some agents play a central role within the network.

Darkly colored nodes are located in the center, indicating that these nodes have a high opinion value. These central agents are likely to influence other agents during the rewiring process.

(2) **Consideration as a Process of Consensus Building**

Many nodes are shown in shades of blue, indicating that many agents have similar opinions. This suggests that opinions are increasingly shared and influenced within the network.

It is observed that some agents have different opinion values compared to other agents. This shows that there is still a diversity of opinions within the network.

(3) **Consideration of the Generation Route of Filter Bubbles**

Many of the nodes located at the center of the network have the same or similar colors, suggesting that there is active exchange of information and opinions between these nodes. This is a typical feature of filter bubbles, showing that agents who share the same opinions and information are strongly connected to each other.

On the other hand, it is observed that nodes located at the periphery of the network have a different color from nodes at the center. This indicates that these agents may have different information sources and opinions than the central agent.

From the above considerations, we can see that the formation of opinions and the propagation of information within a network are greatly influenced by the structure of the network and the relationships between agents. In particular, it is considered that the occurrence of filter bubbles and the polarization of opinions may be strongly influenced by the central agent or group of the network.

This simulation aims to mimic how filter bubbles form within social networks. As discussed in this section, we can hypothesize and verify cases using a sociophysical approach that, due to rewiring processing and restrictions on opinion updating, convergence of opinions among agents and a reduction in diversity, that is, a filter bubble, can be observed.

To simulate a case where the topic is a loop. We set up a conditional equation. It became imperative to modify the rewiring process of the nodes to create a cycle (loop) in the network, and also to adjust the opinion formation process so

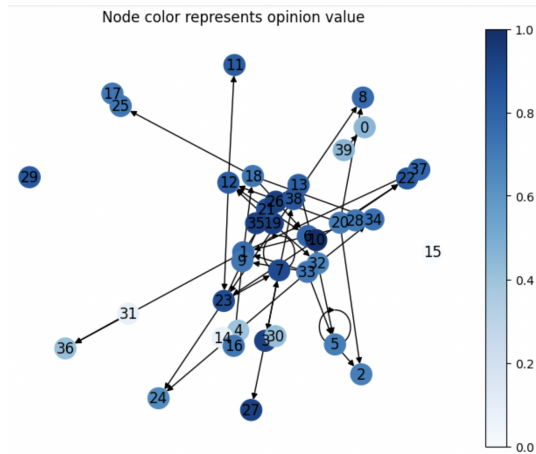

Fig. 3: Reproducing the filter bubble

that the opinion values of the nodes are not updated under certain conditions.

Modified the addition of links in the rewiring process to create a cycle in the network, i.e., the opinion formation process, to stop updating a node's opinion value when it exceeds a certain threshold value.

**Analysis of Network Results**

(1) **Tendency of agent rewiring**

Some nodes (agents) have a large number of ingress and egress links, while other nodes have very few links. This suggests that some agents play a central role within the network.

Darkly colored nodes are located in the center, indicating that these nodes have a high opinion value. These central agents are likely to influence other agents during the rewiring process.

(2) **Consideration as a process of consensus building**

Many nodes are shown in shades of blue, indicating that many agents have similar opinions. This suggests that opinions are increasingly shared and influenced within the network.

It is observed that some agents have different opinion values compared to other agents. This shows that there is still a diversity of opinions within the network.

(3) **Consideration of the generation route of filter bubbles**

Many of the nodes located at the center of the network have the same or similar colors, suggesting that there is active exchange of information and opinions between these nodes. This is a typical feature of filter bubbles, showing that agents who share the same opinions and information are strongly connected to each other.

On the other hand, it is observed that nodes located at the periphery of the network have a different color from nodes at the center. This indicates that these agents may have different information sources and opinions than the central agent.

Multiple cases are constructed in which clusters with initially divergent opinions fall into a filter bubble as multiple clusters are provided with topics. Divide nodes into several clusters and assign different initial values of opinion to nodes in each cluster. Reconstruct connections between nodes based on certain conditions (e.g., a randomly selected node is connected to a node in another cluster). Repeat the convergence of opinions.

**From Figure 4, Discussion**

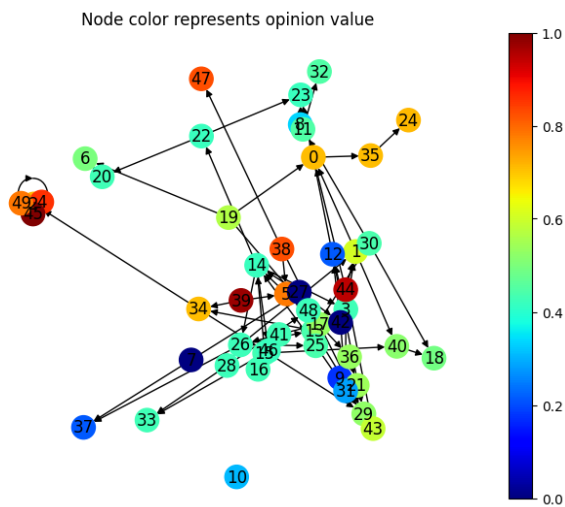

Fig. 4: Fall into a filter bubble as multiple cluster

### 1. Agent rewiring trends

According to the code, 10 rewiring steps are performed. In each step, two nodes from each cluster are randomly selected and the selected nodes are connected to each other.

The color of the graph indicates the opinion value of each node. Initially, the nodes were divided into three different clusters, with each cluster assigned a different initial value of opinion.

Observing the graph, we can see edges between nodes of different colors, indicating that agents with different opinion values are connected to each other during the rewiring process. This suggests that agents may be exposed to diverse opinions.

### 2. Consideration as a process of consensus building

The color gradient indicates convergence or diffusion of opinions among agents. When nodes of different clusters are connected, the opinions of the agents may be attracted to the opinions of the agents to which they are connected.

From the color distribution on the graph, it appears that some agents have intermediate opinion values. This indicates that agents with many connections between different clusters may be exposed to a variety of opinions and thus form neutral opinions.

### 3. What kind of cases and opinion formations can cause this kind of phenomenon in a society?

For example, experiences in international environments and multicultural communities often enrich people's thinking and values. This phenomenon is similar to the rewiring of agents and changes in opinion described above; in SNS and news media, people's opinions and perceptions can change as a result of exposure to diverse sources of information.

### 4. Consideration as a filter bubble phenomenon

Clustering of colors in a graph can be considered as an example of the filter bubble phenomenon. Clusters in which certain opinions and information are concentrated suggest that agents within that group are exposed to similar information and opinions.

Isolated clusters of different colors on the graph mean that each cluster exists within a different bubble of information or opinion.

### 5. Consideration of rewiring route trends

Reroute rewiring serves as a pathway for agents to be exposed to new information and opinions. This can also be a factor that causes an agent's opinion to change.

Observing the distribution of edges in the graph, we see several reroutes between agents with different opinions. This suggests that agents have a chance to be exposed to diverse opinions and information. On the other hand, if there is a dense rewiring within a particular cluster, agents within that cluster are more likely to be exposed primarily to the same information and opinions. This situation can be a factor that reinforces the filter bubble.

**Scenarios where multiple clusters of differing opinions fall into a filter bubble**

When constructing multiple cases where multiple clusters with initially divergent opinions fall into a filter bubble as a result of multiple clusters being provided with topics. Divide the nodes into several clusters and assign different initial values of opinion to the nodes in each cluster. Reconstruct connections between nodes based on certain conditions (e.g., randomly selected nodes are connected to nodes in other clusters). The case of repeated convergence of opinions was verified.

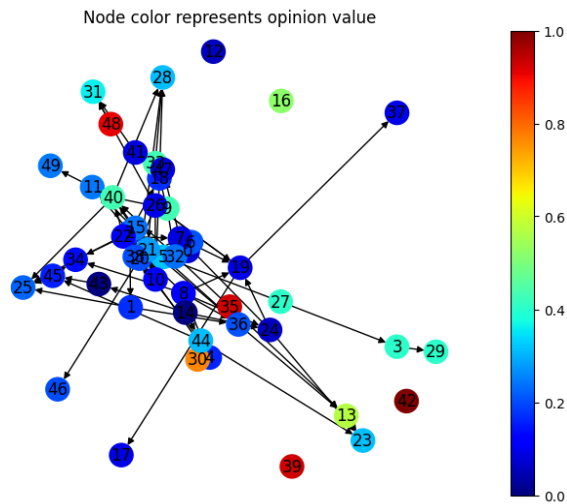

Fig. 5: Fall into a filter bubble as multiple cluster, multiple clusters of differing opinion

(1) **Agent rewiring trends**
Although many connections exist between multiple agents, some agents appear to be rarely rewired.
It is shown that agents of a particular color (opinion value) are frequently connected to agents of other colors. This is evidence of an exchange of opinions among agents with different opinions.

(2) **Consideration as a process of consensus building**
The central clustering of blue colors suggests that a strong consensus of opinion has developed among these agents.
On the other hand, the red and green agents are dispersed and these agents are likely to have different opinion values.

(3) **What is the case in society, opinion formation**
This graph can be said to show strong agreement of opinion within a particular group and diversity of opinion among different groups. In society this could indicate the formation of common values and beliefs within a particular community or group and the diversity of opinions among different groups or communities.

(4) **What is the case for the filter bubble phenomenon?**
An area with a high concentration of blue-colored agents may indicate the presence of a filter bubble of information and opinions. It is likely that these agents are primarily exposed to information from the same or similar sources.

(5) **Root tendency to rewire**
We observe that some agents are rewired with many other agents. This may indicate that these agents play a central role in the exchange of information and opinions.
On the other hand, there are also agents that are less rewired. This may indicate that these agents rely on limited sources of information or that they exchange few opinions with other agents.

**Scenarios in which a cluster diverges the moment a topic provider with a different external force emerges**

In the case of a scenario in which a cluster diverges the moment a topic provider with a different external force emerges after falling into a filter bubble. In the case of multiple cases where clusters with initially divergent opinions fall into a filter bubble when multiple clusters provide topics for discussion. Divide the nodes into several clusters and assign initial values of opinions close to the nodes in each cluster. Based on certain conditions, reconstruct the connections between nodes to fall into the filter bubble. After a certain step, we add a different topic provider as an external force and strongly change the opinion value of that node. We took the hypothesis of repeated convergence of opinions and ran the simulation.

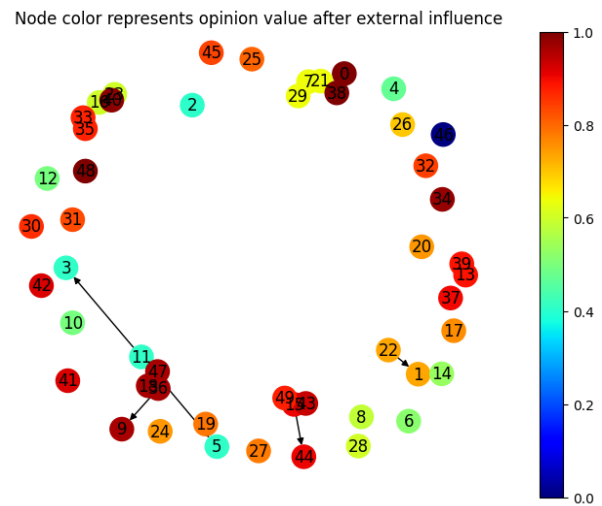

Fig. 6: Fall into a filter bubble as multiple cluster, multiple clusters of differing opinion

(1) **Agent rewiring trends**
The small number of links between nodes suggests relatively little rewiring between agents. There are few indications that agents of a particular color (opinion value) are frequently connected to agents of other colors. This may indicate that the exchange of opinions is limited or takes place only within specific groups.

(2) **Consideration as a process of consensus building**

Red and green agents are dispersed, inferring that these agents are more likely to have different opinion values. On the other hand, agents that are concentrated in a particular area may suggest that there is a strong consensus of opinion within that group.

(3) **What is the case in society, opinion formation**

The graph can be said to indicate a strong agreement of opinion within a particular group and a diversity of opinion among different groups. In society, it may indicate the formation of common values and beliefs within a particular community or group, as well as diversity of opinion among different groups or communities.

(4) **What is the case for the filter bubble phenomenon?**

Areas with high concentrations of blue agents may indicate the presence of filter bubbles of information and opinion. These agents are likely to be primarily exposed to information from the same or similar sources.

(5) **Root tendency to rewire**

It is observed that some agents are rewired with many other agents. This may indicate that these agents play a central role in the exchange of information and opinions. On the other hand, there are some agents that are less rewired. This may indicate that these agents rely on limited sources of information or that they exchange few opinions with other agents.

**Cases in which the filter bubble is exacerbated by the strengthening**

From Figure 7, this cases in which a cluster that was initially connected by close opinions falls into a filter bubble when multiple clusters provide topics, and then the external cluster is removed the moment a topic provider with a different external force is found, and even if the same external force is involved over and over again, the connection of clusters that are even closer is strengthened, and the filter bubble is worsened. We hypothesize a case in which the bubble worsens.

We create clusters, assign each a node with a close opinion value, and after a few steps, allow a strong external opinion provider to enter the network, and remove it from the network immediately after this external cluster joins the network. This repeated entry and removal of this outside opinion provider into the network was set up so that the existing clusters would become stronger and opinions would converge more.

(1) **Tendency of agent rewiring**

Due to the small number of links between agents, there appears to be little rewiring between agents with a particular opinion value (color). There is little indication of frequent connections between agents of different colors. This may indicate that the exchange of opinions is limited or takes place only within specific groups.

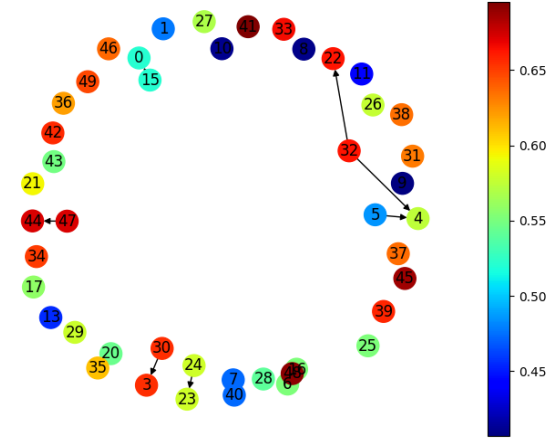

Fig. 7: Cases in which the filter bubble is exacerbated by the strengthening

(2) **Consideration as a process of consensus building**

If agents are concentrated in one particular area, this may suggest that there is a strong consensus of opinion within that group. On the other hand, agents that are spread out are more likely to have different opinion values.

(3) **What is the case in society, opinion formation**

This graph can be said to indicate a strong consensus of opinion within a particular group and a diversity of opinion among different groups. In society, this could indicate the formation of common values and beliefs within a particular community or group and the diversity of opinions among different groups or communities.

(4) **What is the case for the filter bubble phenomenon?**

Areas with a high density of blue agents may indicate the presence of filter bubbles of information and opinion. These agents may be primarily exposed to information from the same or similar sources.

(5) **Reroute trends in rewiring**

It is observed that some agents are rewired with many other agents. This may indicate that these agents play a central role in the exchange of information and opinions. On the other hand, there are also agents that are not or less rewired. This may indicate that these agents rely on limited information sources or have little exchange of ideas with other agents.

**Cases in which the filter bubble bursts**

From Figure 9, after a cluster that was initially connected by close opinions falls into a filter bubble due to multiple clusters providing topics, the external cluster is removed the moment a topic provider with a different external force is introduced, further strengthening the connection of clusters that are close to each other even if the same external force is involved over and over again. However, we will hypothesize and discuss the case in which that filter bubble bursts when more and more opposing clusters with even stronger opinions approach the filter bubbled cluster.

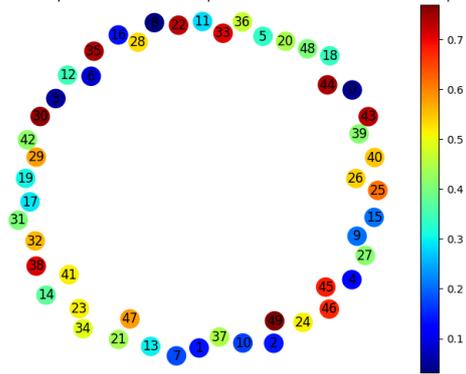

Fig. 8: Cases in which the filter bubble is broken

In this case, an initial node cluster is created to form the filter bubble, At certain steps, a strong external opinion provider enters the network. This external cluster is removed immediately after joining the network, and this process is repeated to assume a pattern in which the opinions of the existing clusters become stronger. Then, when a new rebuttal cluster is created and its opinion is strengthened, the pattern is that the filter bubble bursts when the rebuttal cluster approaches the filter-bubbled cluster.

(1) **Tendency of agents to rewire**

Looking at the distribution of agent colors, it appears that there are few links between agents with specific opinion values. Based on color transitions, we can see that agents of certain colors are concentrated in certain areas.

(2) **Consideration as a process of consensus building**

There are areas where agents of a particular color are densely concentrated. This may suggest a strong consensus of opinion within that group. On the other hand, agents with a spread of colors are more likely to have different opinion values. In other words, opinions are disparate and divergent, reproducing a divergent state of affairs.

(3) **What is the case in society, opinion formation**

This graph can be said to show strong agreement of opinion within a particular group and diversity of opinion among different groups. Socially, it may indicate the formation of common values and beliefs within a particular community or group and the diversity of opinions among different groups or communities. Here, too, a state of disparate and divergent opinions is reproduced.

(4) **What is the case for the filter bubble phenomenon?**

High density areas of blue agents may indicate the presence of filter bubbles of information and opinion. These agents may be receiving information primarily from the same or similar sources. Although partially in conflict with the phenomenon in (4), the overall state is divergent.

(5) **Consideration of root tendencies of rewiring**

We observed that some agents are rewired with many other agents. This may indicate that these agents play a central role in the exchange of information and opinions. On the other hand, some agents are not rewired or not rewired very much. This may indicate that these agents rely on limited information sources or have little exchange of ideas with other agents. We can speculate that these may be the causes of the disparate and divergent opinions that are being reproduced.

**Opposing side becomes a filter bubble**

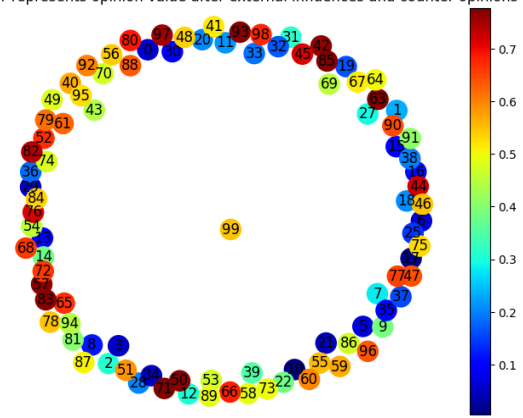

Fig. 9: Opposing side becomes a filter bubble

From Figure 9, after a cluster initially connected by close opinions falls into a filter bubble due to multiple clusters providing topics, the external cluster is removed the moment a topic provider with a different external force appears, and the connection between the clusters that are close is further strengthened even if the same external force is involved over and over again. However, when the number of rebuttal clusters with stronger opinions increases and approaches the filter bubble cluster, the filter bubble side partially encroaches on the rebuttal cluster, and we assume a case in which the rebuttal side is eventually caught up in the filter bubble side.

A filter bubble is formed in the initial node cluster, and strong outside opinion providers continuously enter the network and are immediately removed, creating a counter-argument cluster, which strengthens its opinion. Then, when the refuting cluster approaches the filter-bubbled cluster, the case is constructed in which the partially refuting cluster is eroded to the filter-bubble side, and eventually the refuting party is also caught up in the filter-bubble side.

(1) **Agent rewiring trends**
Areas of relatively high density of nodes (especially in the range of 0.6 to 0.7) are observed from red to blue. This indicates that agents with these opinion values may be strongly connected to each other. On the other hand, nodes in the 0.1 to 0.3 range are dispersed throughout the graph, suggesting less rewiring among these agents.

(2) **Consideration as a process of consensus building**
Areas with a high density of agents of a darker color (0.6 or higher) are more likely to have formed a consensus of opinion within that group. On the other hand, areas with a mix of agents of different colors indicate that a diversity of opinions exists.

(3) **What is the case within a society, opinion formation**
We can say that this graph shows a strong agreement of opinions within a particular group and a diversity of opinions among different groups. Socially, it may indicate the formation of common values and beliefs within a particular community or group, and diversity of opinion among different groups or communities.

(4) **What is the case for the filter bubble phenomenon?**
Dense areas of dark-colored agents may indicate the presence of filter bubbles of information and opinion. These agents may be receiving information primarily from the same or similar sources.

(5) **Consideration of rewiring root tendencies**
We observed agents that are rerouted with many other agents. This indicates that these agents may play a central role in the exchange of information and opinions. On the other hand, there are also agents that are not rewired or not very rewired. This may indicate that these agents rely on limited information sources or have little exchange of ideas with other agents.

**Case in which the third opinion becomes mainstream**

From Case of Figure 11, after a cluster initially connected by close opinions falls into a filter bubble due to multiple clusters providing topics, the external cluster is removed the moment a topic provider with a different external force is found, further strengthening the connection of the close clusters even if the same external force is involved over and over again. However, the filter bubble side partially encroaches on the opposing clusters when more and more opposing clusters with stronger opinions approach the filter-bubbled cluster.

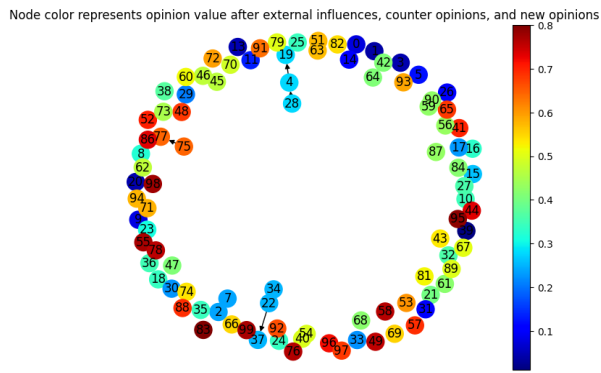

Fig. 10: Case in which the third opinion becomes mainstream

However, a new opinion is generated here, and a new third opinion is generated from the two clusters, and we assume a case in which that cluster eventually becomes stronger. A filter bubble is formed in the initial node cluster, and outside opinion providers continuously enter the network and are removed shortly after. At that point, a counter-opinion cluster is generated, its opinion becomes stronger, and as the counter-opinion cluster approaches the filter-bubbled cluster, the counter-opinion cluster partially erodes to the filter-bubble side. Assume a case where a new opinion cluster is generated and it eventually becomes the most influential cluster.

**When two different opinions are strong**

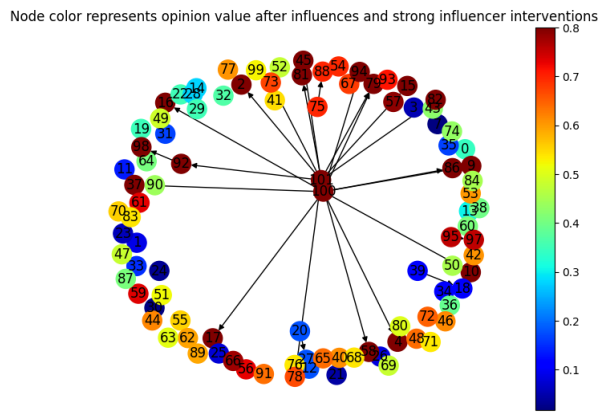

Fig. 11: When two different opinions are strong

From Case of Figure 11, after a cluster that was initially connected by close opinions falls into a filter bubble due to multiple clusters providing topics, the external cluster is removed the moment a topic provider with a different external force is found, further strengthening the connection of the clusters that are close even if the same external force is involved over and over again. However, assume a case where the intervention of two influential people of a certain opinion pulls in these two different opinions at once. A filter bubble is formed in the initial node cluster, and outside opinion providers continuously enter the network, only to be removed shortly after. Then two people with strong opinion influence enter the network and pull the network's opinions in two different directions. Assume a case where some nodes remain separated from the opinions of the two people.

(1) **Agent rewiring trends**

It is clear that the agent in the center of the graph (node indicated as 10) is directly connected to many other agents. This indicates that this agent is very influential or plays a central role.

Based on color, we see that agents of diverse opinions are connected to the central agent. This confirms that the central agent may have access to diverse sources of information and that the hypothetical case can be reproduced by the model.

(2) **Consideration as a process of consensus building**

The dark-colored agents (0.6 and above) are relatively evenly distributed, but the mixed colors of the agents connected to the central agent suggest that consensus may be in progress or that there is an active exchange of different opinions.

(3) **What is the case in society, opinion formation**

The graph may indicate a scenario where there is one central source or leader who has a direct relationship with a number of individuals or groups. For example, it might reflect a situation such as corporate, organizational, or community leadership.

(4) **What is the case for the filter bubble phenomenon**

Because the central agent is directly connected to many other agents, the information this agent receives may be diverse. However, if an external agent receives information only through the central agent, it may facilitate the formation of filter bubbles.

(5) **Consideration of rewiring root tendencies**:

The central agent has the most connections in the network and is likely to be the most frequently rerouted agent. It is assumed that other agents often exchange opinions and information via the central agent.

**Concentration of Central Opinions**

From Case of Figure 12, After a cluster that was initially closely connected by opinion falls into a filter bubble due to multiple clusters providing topics, the external cluster is removed the moment a topic provider with a different external force becomes involved, further strengthening the connection of the clusters that are close, even if the same external force is involved over and over again. However, the two most influential people of a given opinion repeatedly change their opinions. This repetition leads to frequent changes in the surrounding opinions, and the final case is that most opinions fall apart, while some are pulled by the repeated opinions of the two most influential people.

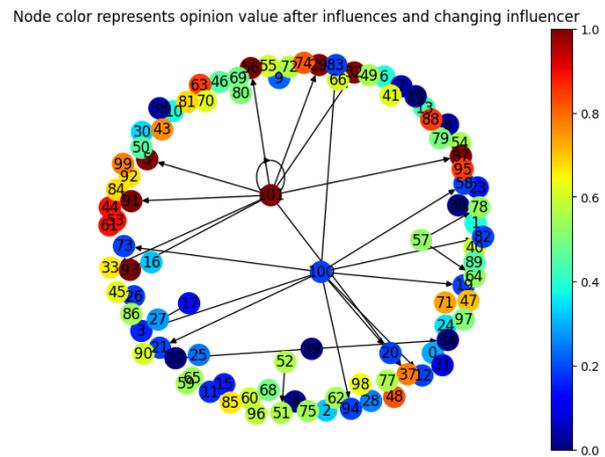

Fig. 12: Concentration of Central Opinions

(1) Agent rewiring trends

The agent located in the center of the graph (node 10) is directly connected to many other agents. This suggests that this agent may be very influential or play a central role.

Based on color, agents of varying opinions are connected to the central agents. This confirms that the two central agents may have access to diverse sources of information.

(2) Consideration as a process of consensus building

While the dark agents (0.6 and above) are relatively evenly distributed, the mixed colors of the agents connected to the central agent suggest that a consensus is in progress or that there is an active exchange of different opinions.

(3) What is the case in society, opinion formation

The graph may indicate a scenario in which the two central sources or leaders have direct relationships with a large number of individuals or groups. For example, it may reflect a situation such as corporate, organizational, or community leadership.

(4) What is the case for the filter bubble phenomenon?

Since the central agents are directly connected to many other agents, the information this agent receives could be diverse. However, if outside agents receive information only through the central agents, this may facilitate the formation of filter bubbles.

(5) Consideration of Root Tendency of Rerouting

The central agents are the most connected in the network and are likely to be the most frequently rerouted agents. Other agents are presumed to frequently exchange ideas and information via the central two agents.

## 4. Filter bubble on Propagation Probability in Network Structures

We now turn to a discussion of propagation probabilities in simulations of filter bubble generation. Propagation probability serves as a quantitative measure that captures the characteristics of information flow and connectivity within a network. Based on the likelihood of information reaching another node from a particular node, this probability is calculated as follows

$$P_{\text{propagation}}(i \rightarrow j) = \frac{\text{Number of reachable nodes from } i}{\text{Total number of nodes}} \quad (5)$$

Here, $P_{\text{propagation}}(i \rightarrow j)$ represents the propagation probability from node $i$ to node $j$. This probability is highly dependent on the topology of the network and the connectivity of each node.

The connectivity within the network reflects the efficiency of overall information flow and the degree of interactions among nodes within the network. A high propagation probability indicates a smooth flow of information across the network, while a low probability suggests a tendency for information to remain within certain portions.

The calculation of propagation probability involves a process of assessing the reachability of each node within the network. This is conducted following these steps:

(1) Identify the shortest paths from each node $i$ to all other nodes within the network.
(2) Count the number of nodes reachable from each node $i$.
(3) Calculate the average number of reachable nodes across all nodes, normalizing this by the size of the network.

**Discussion on the possibility of Propagation Probability**

From Figure 13 and Figure 14, discussion on the possibility of Propagation Probability,

(1) **Trend of Agent Rewirings**
The "Distribution of Rewirings per Node" graph shows

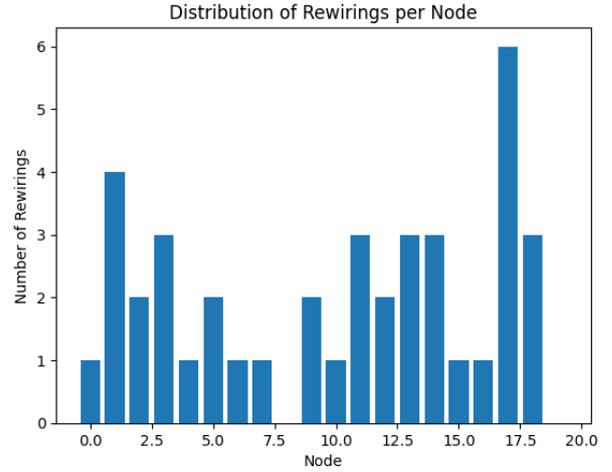

Fig. 13: Distribution of Rewirings per Node $N = 20$

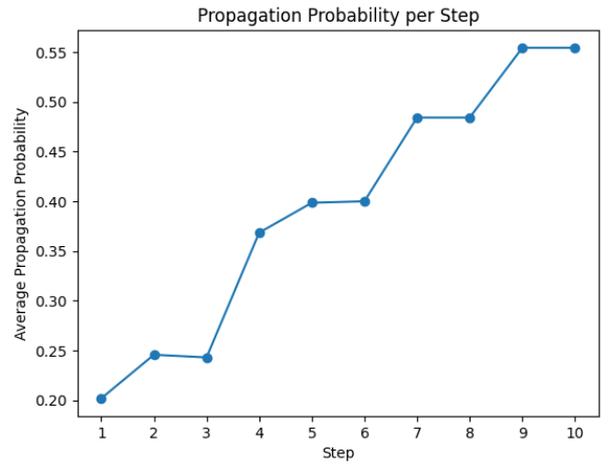

Fig. 14: Propagation Probability per Step $N = 20$

that many rewirings are performed at certain nodes. In particular, we can see that there are a great many rewirings in the vicinity of node 20. This indicates that the node plays a central role in the network or that rewiring occurs frequently due to some other factor.

(2) **Consideration as a consensus building process**
The "Propagation Probability per Step" graph shows that the propagation probability increases with each step of rewiring. This indicates that information propagation in the network is becoming more efficient as rewiring progresses. It can be confirmed that when each agent can receive information more easily, the possibility of smooth progress in consensus building will increase.

(3) **What is the case in society and opinion formation**
This indicator may indicate the influence of a leader or central figure in a society on information transmission and opinion formation. If a particular agent plays a central role, we can see if that agent's opinions and information are more likely to influence other agents.

(4) **What is the case for the filter bubble phenomenon?**
If a central agent communicates only consistent opinions and information, other agents in the network are more likely to be influenced by those opinions and information. This could lead to the formation of a filter bubble, and parameters could be checked, keeping in mind that diverse opinions and information may be less likely to be propagated in the network.

(5) **Consider the rewiring routing trends.**
A closer look at the code shows that the probability of a node with an opinion value of 0.5 or higher being selected for rerouting is decreasing. This indicates that rerouting routes are adjusted based on opinion values. Also, rerouting is done primarily between the four selected nodes. This may strengthen the connections between certain agents in the network. Overall, the code and graph illustrate the process of information propagation, association among agents, and opinion formation within the network. It suggests that when a particular agent plays a central role, its impact on the network as a whole can be considered.

## 4.1 Stubbornness Probability

This approach allows for capturing the dynamics of information propagation under specific scenarios, assessing the impact of network structure on information propagation.

The calculation of stubbornness probability is based on the idea that agents are considered stubborn if their opinion values exceed a certain threshold, countering the propagation trend. This is represented by the following equation:

$$P_{\text{stubborn}} = 1 - \frac{1}{N} \sum_{i=1}^{N} \mathbf{1}_{\{o_i > T\}} \quad (6)$$

where:

$P_{\text{stubborn}}$ represents the probability of stubbornness, indicating the proportion of nodes in the network that are resisting change.

$N$ is the number of nodes in the network.

$\mathbf{1}_{\{o_i > T\}}$ is the indicator function, which is 1 if the opinion $o_i$ of node $i$ exceeds the threshold $T$, and 0 otherwise.

From Figure 15 and Figure 16, discussion on the possibility of Stubbornness Probability,

(1) **Agent rewiring trends**
From the initial diagram of the network, many nodes are connected to a central node. In particular, node 0 has many input edges. Nodes with opinion values greater than 0.5 have a 50% chance of being rewired. This allows us to observe a tendency for certain nodes to be rewired frequently.

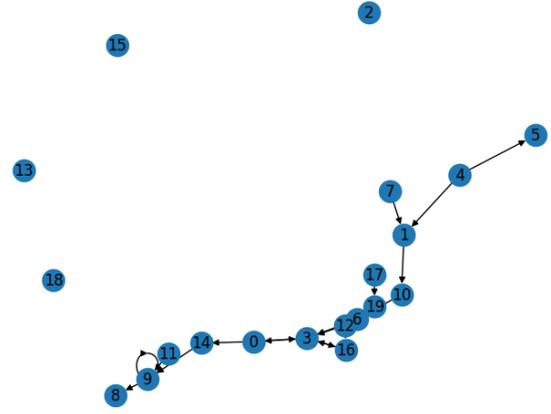

Fig. 15: Stubbornness Probability $N = 20$

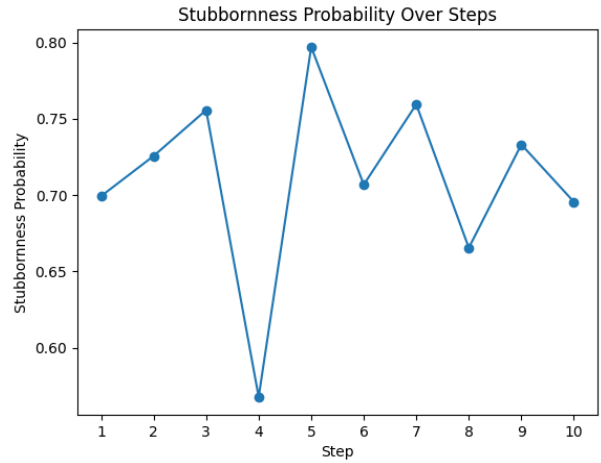

Fig. 16: Stubbornness Probability Over Steps $N = 20$

(2) **Consideration as a process of consensus building**
According to the "Stubbornness Probability Over Steps" graph, the probability of stubbornness at each step is variable. This can affect the efficiency of information propagation and consensus building within the network. As the propagation probability increases, we can observe that the process of consensus building may proceed more smoothly because each agent can receive information more easily.

(3) **What is the case in society, opinion formation**
When an agent plays a central role in a network, there is a greater likelihood that his/her opinions and information will influence other agents. We can envision a simulation hypothesis that can be thought of as an indicator of how the influence of a leader or central figure in a society affects information transfer and opinion formation.

(4) **What is the case for the filter bubble phenomenon?**
When a central agent conveys only consistent opinions or information, other agents in the network may be influenced by that opinion or information. This may lead to the formation of a filter bubble, making it difficult for diverse opinions and information to propagate within the network.

(5) **Consider the root tendency of rewiring.**
It is shown that nodes with opinion values of 0.5 or higher have a decreasing probability of being selected for rerouting. This indicates that the rerouting routes are adjusted based on opinion values. In addition, the rerouting is mainly done between the four selected nodes. This confirms that connections between certain agents in the network may be strengthened.

## 4.2 Propagation probability trends

This calculation allows for the measurement of the tendency towards stubbornness in the entire network after each step, acting inversely to the propagation probability trends.

Given a graph $G$ with nodes $N$ and a set of opinions $O$ where each $o_n$ corresponds to the opinion of node $n$, and each node's opinion is influenced by its predecessors. The influence factor $\alpha$ represents the extent to which a node is affected by the opinions of its neighboring nodes. The updated opinion of a node is calculated as follows:

$$o_n^{\text{new}} = (1 - \alpha) \cdot o_n + \alpha \cdot \left( \frac{\sum_{i \in \text{pred}(n)} o_i}{|\text{pred}(n)|} \right) \quad (7)$$

where:

$o_n^{\text{new}}$ is the new opinion of node $n$,

$\alpha$ is the influence factor dictating the degree of influence,

$\text{pred}(n)$ denotes the predecessors of node $n$,

$o_i$ represents the opinion of the $i$-th predecessor node,

$|\text{pred}(n)|$ is the number of predecessors of node $n$.

This equation ensures that each node's opinion is adjusted by considering a weighted average of its own opinion and the opinions of its predecessors.

# 5. Modeling Opinion Dynamics and Network Topology Evolution in Social Networks

In this study case of social networks, understanding how opinions spread and evolve is crucial. This document outlines a simulation model that combines opinion dynamics with the evolution of the network topology, specifically focusing on how these aspects influence the distribution of opinions (node density) across the network in each step.

## 5.1 Model Description

The model is designed to simulate a social network using directed graphs, where nodes represent individuals, and edges represent the influence between them. Opinions are numerical values assigned to each node, and the network topology evolves through a process of rewiring, influenced by the opinions of the nodes.

## 5.2 Parameters

The main parameters governing the model are as follows:

**N**: The number of nodes in the network.

**rewiring_steps**: The number of steps in the simulation during which edges may be rewired.

**rewire_prob**: The probability of an edge being rewired in each step.

**influence_factor**: A factor representing how much a node's opinion is influenced by its neighbors.

### 5.2.1 Network Initialization and Edge Creation

The network is initialized with $N$ nodes and a set number of edges created randomly between them. The *create_random_edges* function handles the initial edge creation, ensuring a randomly generated network topology.

$$\text{create\_random\_edges}(G, N) : G(V, E) \to G(V, E') \quad (8)$$

where $G$ is the graph representing the network, $V$ is the set of nodes, $E$ and $E'$ are the sets of edges before and after the function execution, respectively.

### 5.2.2 Opinion Update

Each node's opinion gets updated based on the opinions of its neighbors. This mechanism is represented by the *update_opinions* function.

$$\begin{aligned}\text{new\_opinions}[i] = &\,(1 - \text{influence\_factor}) \cdot \text{opinions}[i] \\ &+ \left( \frac{\sum(\text{opinions}[j])}{|N(i)|} \right) \cdot \text{influence\_factor}\end{aligned} \quad (9)$$

where $i$ indexes the current node, $j$ indexes the neighbors of node $i$, $N(i)$ represents the set of neighbors, and opinions represents the opinion values.

## 5.3 Edge Rewiring

Edges are rewired based on the *rewire_edges* function, which depends on the nodes' opinions and the rewiring probability.

rewire_edges : $G(V, E) \to G(V, E'')$,

where $G$ is the graph,

$V$ is the set of nodes,

$E$ and $E''$ are the sets of edges before and after rewiring.

where $E''$ represents the set of edges after possible rewiring.

### 5.3.1 Opinion Distribution Calculation

The distribution of opinions (node density) within the network is calculated using a histogram method, which divides the range of opinions into bins and counts the number of nodes with opinions within each bin.

$$\text{calculate\_opinion\_density}(G, \text{opinions}) : \\ G(V) \times \text{opinions} \to \text{density} \quad (10)$$

where density is the normalized count of nodes for each opinion bin, representing the probability density of the opinions across the network.

## 5.4 Calculation of Opinion Distribution Change Rate

In the analysis of opinion dynamics, understanding the rate of change in opinion distribution is crucial as it provides insights into the volatility or stability of opinion formation within the network. This section describes the mathematical computation used to ascertain the change rate in opinion distribution across consecutive steps in the simulation.

### 5.4.1 Change Rate Formula

The change rate of opinion distribution between two consecutive steps is calculated using the formula:

$$\text{change\_rate} = \frac{\text{current\_density} - \text{previous\_density}}{\text{previous\_density} + \epsilon} \quad (11)$$

where:

current_density is the opinion distribution in the current step.

previous_density is the opinion distribution in the previous step.

$\epsilon$ is a very small number (e.g., $1e-10$) to prevent division by zero.

### 5.4.2 Parameters and Variables

The change rate calculation uses the following parameters and variables:

**current_density** The array representing the density of opinions across different bins in the current step, computed as the number of nodes in each opinion bin divided by the total number of nodes.

**previous_density** The array representing the density of opinions across different bins in the previous step, computed in the same manner as current_density.

**epsilon** A small constant to prevent division by zero during the computation of the change rate. This is necessary because the previous density could be zero for some opinion bins, making the denominator zero.

## 5.5 Calculation of the Probability of Unchanged Opinions

The change rate indicates the relative change in the opinion distribution of the network. Positive values represent an increase in the density of certain opinions, while negative values indicate a decrease. Analyzing these change rates helps in understanding the dynamics and possibly predicting trends in opinion changes. In the network opinion dynamics simulation, one of the metrics of interest is the probability that opinions do not change between consecutive steps. This document outlines the formula used to calculate this probability and describes the parameters involved.

**Probability of Unchanged Opinions**

The probability of unchanged opinions, referred to as the static opinion probability, is computed at each step, considering the change rates in opinion distribution and certain probabilities defining agent behaviors. The formula for the static opinion probability is given by:

$$P_{\text{static}} = (1 - |R_{\text{change}}|) \cdot (1 - P_{\text{prop}}) \cdot (1 - P_{\text{stubborn}}) \cdot (1 - P_{\text{update}}) \quad (12)$$

where each term is defined as follows:

$P_{\text{static}}$: The probability of an opinion remaining unchanged in the current step.

$R_{\text{change}}$: The rate of change in opinion distribution from the previous step to the current step, calculated as the difference in opinion densities divided by the sum of the opinion densities plus a small constant ($\varepsilon$) to prevent division by zero.

$P_{\text{prop}}$: The probability of an opinion propagating from one agent to another. It reflects the likelihood that an agent's opinion will be adopted by others in the network.

$P_{\text{stubborn}}$: The probability associated with agents maintaining their current opinions regardless of influence from their neighbors. It represents the level of stubbornness among agents in the network.

$P_{\text{update}}$: The probability of an agent updating its opinion based on influences within the network. This parameter controls how frequently agents reconsider and potentially change their opinions.

The product in the formula represents the multiplicative effect of various factors that could contribute to an opinion remaining unchanged. A higher rate of change ($R_{\text{change}}$) implies a more dynamic opinion landscape where changes are more frequent, thereby reducing the probability of static opinions. Conversely, higher propagation, stubbornness, or update probabilities contribute to maintaining the status quo, increasing the likelihood that opinions remain static during each step of the simulation.

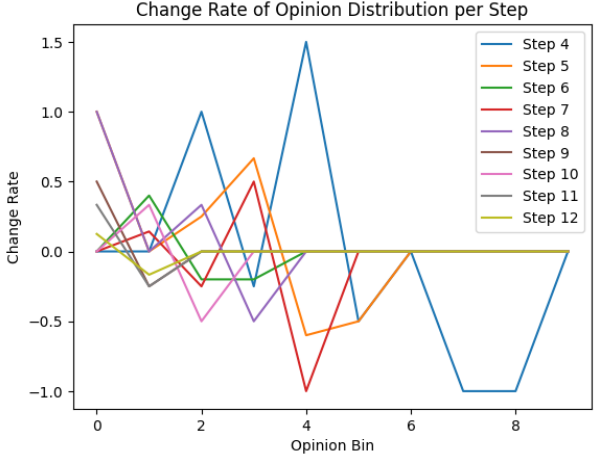

Fig. 17: Change Rate of Opinion Distribution per Step $N = 20$

From Figure 17 and Figure 18, discussion on the possibility of Opinions by Steps value,

(1) **Agent rewiring propensity**

The rewiring probability (`rewire_prob`) is set at 0.5. This indicates that edges between nodes with opinion differences greater than 0.5 have half the probability of being rewired. In other words, it implies that a pair of agents with a large difference of opinion has a 50% reduced probability of being connected to each other.

(2) **Consideration as a process of consensus building**

From the graph, we can see that the rate of change in opinion increases at several steps. In particular, the sharp increase or decrease in the rate of change in several Opinion Bins suggests that convergence or dispersion of opinions may be underway. With a `stubbornness_prob` of

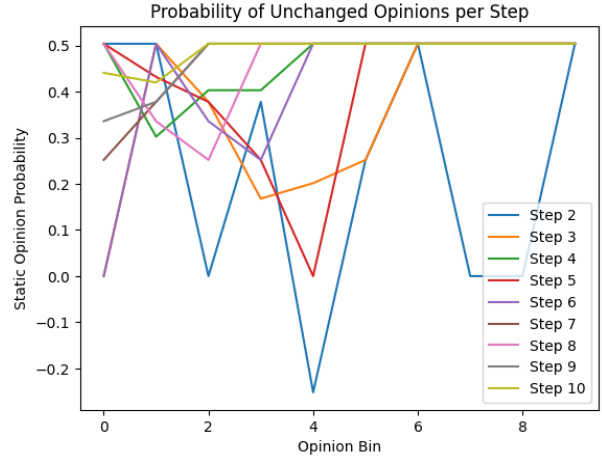

Fig. 18: Probability of Unchanged Opinions per Step $N = 20$

0.2, agents have a 20% chance of sticking to their opinions. This can suggest that it may be difficult to reach a complete consensus.

(3) **What is the case in society and opinion formation**

The consequences of this case are appropriate for situations where people tend to break off relations with those who hold strongly different opinions. For example, the results suggest that this may be useful when considering the formation or change of opinions on topics that people feel strongly about, such as political opinions or religious beliefs.

(4) **What is the case for the filter bubble phenomenon?**

A filter bubble is a phenomenon in which individuals share and exchange information only with those who share the same opinions and ideas, thereby reinforcing certain opinions and beliefs. In this model, `rewire_prob` rewires edges between agents with different opinions, suggesting the possibility of the filter bubble phenomenon.

(5) **Please consider the root tendency of rewiring**

The code states that edges between nodes with opinion differences greater than 0.5 may be rewired. This rewiring trend suggests that agents with large gaps in opinion may break connections and form new connections between agents with close opinions. This suggests that clusters of agents with the same opinions are more likely to form.

**Rewire and Calculate Function**

The function `rewire_and_calculate` contains probabilistic operations based on random numbers to decide on removing or adding edges to the nodes in the network.

(1) **Edge Removal:** An edge between a node and one of its neighbors is removed based on a probability. The corresponding conditional expression is:

$$P(\text{remove\_edge}) = \begin{cases} 1 & \text{if rand()} < p \\ 0 & \text{otherwise} \end{cases} \quad (13)$$

Explanation: An edge from the current node to a randomly selected neighbor is removed with a probability of $p$.

(2) **Edge Addition:** An edge is added between a node and a non-neighbor node based on a probability. The corresponding conditional expression is:

$$P(\text{add\_edge}) = \begin{cases} 1 & \text{if rand()} < q \\ 0 & \text{otherwise} \end{cases} \quad (14)$$

Explanation: An edge between the current node and a randomly selected non-neighbor node is added with a probability of $q$.

(3) **Network Density:** The density of the network is calculated using:

$$\text{density} = \frac{\text{number of actual edges}}{\text{number of potential edges}} \quad (15)$$

Explanation: Network density gives the ratio of the actual number of edges to the number of potential edges in the graph.

(4) **Propagation Probability:** The propagation probability is randomly generated in the given range:

$$\text{propagation\_probability} = \text{rand()} \times a + b \quad (16)$$

Explanation: This is a placeholder for the propagation probability. The value is generated randomly between $a$ and $b$.

**Network Density and Propagation Probability Over Steps**

The function appends the calculated density and propagation probability for each rewiring step. These values are then plotted over steps to visualize the changes. The relevant expressions from the code are:

$$\text{densities}[i] = \text{density at step } i \quad (17)$$
$$\text{propagation\_probabilities}[i] = \text{propagation probability at step } i \quad (18)$$

Explanation: For each step in the rewiring process, the density of the network and the propagation probability are recorded. These are then used for plotting.

**Correlation Between Density and Propagation Probability**

The scatter plot illustrates the relationship between the network density and propagation probability. Each point represents a rewiring step. The relevant expressions are:

$$(x_i, y_i) = (\text{density at step } i, \text{propagation probability at step } i) \quad (19)$$

Explanation: Each point in the scatter plot represents the density and propagation probability of the network at a particular rewiring step.

## 6. Disucussion

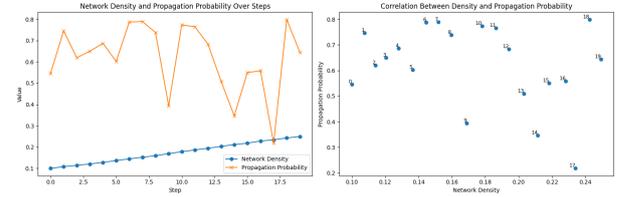

Fig. 19: Network Density and Propagation Probability Over Steps and Correlation Between Density and Propagation Probability $N = 20$

From Figure 19, discussion Density and Propagation Probability by Steps value,

(1) **Agent rewiring trends**
The graph on the left shows that the density of the network is consistently decreasing over time. On the other hand, the propagation probability is very variable, but generally high, with large increases and decreases at specific steps. This can suggest that agents are more likely to break off relationships with other agents and form relationships with new agents.

(2) **Consideration as a process of consensus formation**
Variations in propagation probability may indicate the process of opinion propagation and consensus formation among agents. A sudden increase or decrease in propagation probability at a particular step may suggest that convergence or dispersion of opinions is underway.

(3) **What is the case in society, opinion formation**
In a society, this case may apply when people tend to disassociate themselves from those with strongly differing opinions. The results could be useful and suggestive when considering the formation and change of opinions on topics that people feel strongly about, such as political opinions or religious beliefs.

(4) **What is the case for the filter bubble phenomenon?**
A filter bubble refers to a phenomenon in which information is shared only among people who share the same opinions and ideas, reinforcing certain opinions and beliefs. The relationship between the data points in the graph on the right suggests the possibility of rewiring of relationships between agents with different opinions, i.e., the filter bubble phenomenon.

(5) **Consider the root tendency of rewiring**
The consistent decrease in network density shown in the left graph may indicate a tendency for connections between agents with widely differing opinions to be broken and for new connections to form between agents with the same opinions. This indicates that clusters of agents with the same opinions are likely to form.

# 7. Conclusion, Perspect

This study presents a theory of opinion dynamics that considers each person's opinion as a continuous value rather than a discrete value. Opinions are represented as real numbers ranging from positive to negative. Trust and distrust are introduced as coefficients for each person pair. A mathematical model was constructed that incorporates external pressure in addition to the influence of opinion exchange within each group. By using this theory, we aim to formulate hypotheses and mathematically represent many phenomena that can occur in group societies.

The filter bubble case, which is based on the non-consensus-based opinion dynamics theory in this paper, allows us to compute the dynamics of a complex system in which people have a mixture of trust and doubt. It can also account for situations in which opinions become increasingly radical because there is no upper limit to opinions. Simulations of large numbers of people are also available. In the future, we will compare and examine whether this theory is consistent with data on speech in actual political and social issues, and what cases we expect to see. In the future, we will compare and examine what cases we assume this theory is consistent with actual data on speech in actual political and social issues.

# Aknowlegement

This research is supported by Grant-in-Aid for Scientific Research Project FY 2019-2021, Research Project/Area No. 19K04881, "Construction of a new theory of opinion dynamics that can describe the real picture of society by introducing trust and distrust".

It is with great regret that we regret to inform you that the leader of this research project, Prof. Akira Ishii, passed away suddenly in the last term of the project(2021). Prof. Ishii was about to retire from Tottori University, where he was affiliated with at the time.

However, he had just presented a new basis in international social physics, complex systems science, and opinion dynamics, and his activities after his retirement were highly anticipated. It is with great regret that we inform you that we have to leave the laboratory. We would like to express our sincere gratitude to all the professors who gave me tremendous support and advice when We encountered major difficulties in the management of the laboratory at that time.

First, Prof. Isamu Okada of Soka University provided valuable comments and suggestions on the formulation of the three-party opinion model in the model of Dr. Nozomi Okano's (FY2022) doctoral dissertation. Prof.Okada also gave us specific suggestions and instructions on the mean-field approximation formula for the three-party opinion model(Equation (13)), Prof.Okada's views on the model formula for the social connection rate in consensus building, and his analytical method. We would also like to express our sincere gratitude for your valuable comments on the simulation of time convergence and divergence in the initial conditions of the above model equation, as well as for your many words of encouragement and emotional support to our laboratory.

We would also like to thank Prof.Masaru Furukawa of Tottori University, who coordinated the late Prof.Akira Ishii's laboratory until FY2022, and gave us many valuable comments as an expert in magnetized plasma and positron research.

In particular, we would like to thank Prof.Hidehiro Matsumoto of Media Science Institute, Digital Hollywood University. Prof.Hidehiro Matsumoto is Co-author of our paper("Case Study On Opinion Dynamics In Dyadic Risk On Possibilities Of Trust-Distrust Model."), for managing the laboratory and guiding us in the absence of the main researcher, and for his guidance on the elements of the final research that were excessive or insufficient with Prof.Masaru Furukawa.

And in particular, Prof.Masaru Furukawa of Tottori University, who is an expert in theoretical and simulation research on physics and mathematics of continuum with a focus on magnetized plasma, gave us valuable opinions from a new perspective.

His research topics include irregular and perturbed magnetic fields, MHD wave motion and stability in non-uniform plasmas including shear flow, the boundary layer problem in magnetized plasmas, and pseudo-annealing of MHD equilibria with magnetic islands.

We received many comments on our research from new perspectives and suggestions for future research. We believe that Prof.Furukawa's guidance provided us with future challenges and perspectives for this research, which stopped halfway through. We would like to express sincere gratitude to him.

We would like to express my sincere gratitude to M Data Corporation, Prof.Koki Uchiyama of Hotlink Corporation, Prof.Narihiko Yoshida, President of Hit Contents Research Institute, Professor of Digital Hollywood University Graduate School, Hidehiko Oguchi of Perspective Media, Inc. for his valuable views from a political science perspective. And Kosuke Kurokawa of M Data Corporation for his support and comments on our research environment over a long period of time. We would like to express our gratitude to Hidehiko Oguchi of Perspective Media, Inc. for his valuable views from the perspective of political science, as well as for his hints and suggestions on how to build opinion dynamics.

We are also grateful to Prof.Masaru Nishikawa of Tsuda Uni-versity for his expert opinion on the definition of conditions in international electoral simulations.

We would also like to thank all the Professors of the Faculty of Engineering, Tottori University. And Prof.Takayuki Mizuno of the National Institute of Informatics, Prof.Fujio Toriumi of the University of Tokyo, Prof.Kazutoshi Sasahara of the Tokyo Institute of Technology, Prof.Makoto Mizuno of Meiji University, Prof.Kaoru Endo of Gakushuin University, and Prof.Yuki Yasuda of Kansai University for taking over and supporting the Society for Computational Social Sciences, which the late Prof.Akira Ishii organized, and for their many concerns for the laboratory's operation. We would also like to thank Prof.Takuju Zen of Kochi University of Technology and Prof.Serge Galam of the Institut d'Etudes Politiques de Paris for inviting me to write this paper and the professors provided many suggestions regarding this long-term our research projects.

We also hope to contribute to their further activities and the development of this field. In addition, we would like to express our sincere gratitude to Prof.Sasaki Research Teams for his heartfelt understanding, support, and advice on the content of our research, and for continuing our discussions at a time when the very survival of the research project itself is in jeopardy due to the sudden death of the project leader.

We would also like to express our sincere gratitude to the bereaved Family of Prof.Akira Ishii, who passed away unexpectedly, for their support and comments leading up to the writing of this report. We would like to close this paper with my best wishes for the repose of the soul of Prof.Akira Ishii, the contribution of his research results to society, the development of ongoing basic research and the connection of research results, and the understanding of this research project.